%% file: Z4Leptons2012Paper.tex
\documentclass[
aps,
prl,
reprint,
showpacs,
]{revtex4-1}
\pdfoutput=1

\usepackage{booktabs}
\usepackage{graphicx}
\usepackage{preprintcover}
\usepackage{multirow}
\usepackage{subfigure}
\usepackage[hyperindex,breaklinks,hidelinks,colorlinks,allcolors=blue]{hyperref}

\newcommand{\xsec}{cross section}
\newcommand{\xsecs}{cross sections}

\def\TeV{\ifmmode {\mathrm{\ Te\kern -0.1em V}}\else
                   \textrm{Te\kern -0.1em V}\fi}
\def\GeV{\ifmmode {\mathrm{\ Ge\kern -0.1em V}}\else
                   \textrm{Ge\kern -0.1em V}\fi}

\newcommand{\W}{\ensuremath{W}}
\newcommand{\Z}{\ensuremath{Z}}

\newcommand{\Zzero}{\ensuremath{Z}}
\newcommand{\WZ}{\ensuremath{\W\Zzero}}

\def\pt{\ensuremath{p_{\mathrm{T}}}}
\def\ET{\ensuremath{E_{\mathrm{T}}}}
\def\GeV{\ifmmode {\mathrm{\ Ge\kern -0.1em V}}\else
                   \textrm{Ge\kern -0.1em V}\fi}%

\def\progname#1{{\sc #1}}
\newcommand{\SingleZllll}{\ensuremath{Z\rightarrow 4\ell}}
\newcommand{\Zllll}{\SingleZllll}

\newcommand{\ZllllFrac}{\ensuremath{\Gamma_{\SingleZllll{}}/\Gamma_{Z}}}
\newcommand{\ZllllFracDisp}{\ensuremath{\frac{\Gamma_{\SingleZllll{}}}{\Gamma_{Z}}}}
\newcommand{\ZmmFrac}{\ensuremath{\Gamma_{Z\rightarrow\mu\mu}/\Gamma_{Z}}}
\newcommand{\ZmmFracDisp}{\ensuremath{\frac{\Gamma_{Z\rightarrow\mu\mu}}{\Gamma_{Z}}}}
\newcommand{\Cllll}{\ensuremath{C_{4\ell}}}
\newcommand{\Allll}{\ensuremath{A_{4\ell}}}
\newcommand{\CAmumu}{\ensuremath{(C \times A)_{2\mu}}}
\newcommand{\SigFid}{\ensuremath{\sigma^{\text{fid}}_{Z4\ell}}}
\newcommand{\SigPS}{\ensuremath{\sigma_{Z4\ell}}}
\newcommand{\NObs}{\ensuremath{N_{4\ell}^{\text{obs}}}}
\newcommand{\NExp}{\ensuremath{N_{4\ell}^{\text{exp}}}}
\newcommand{\NBkg}{\ensuremath{N_{4\ell}^{\text{bkg}}}}
\newcommand{\NObsMuMu}{\ensuremath{N_{2\mu}^{\text{obs}}}}
\newcommand{\NBkgMuMu}{\ensuremath{N_{2\mu}^{\text{bkg}}}}

\input{Numbers.tex}

\PreprintCoverPaperTitle{\protect\input{00_Title.tex}}

\PreprintIdNumber{CERN-PH-EP-2014-047}

\PreprintJournalName{Phys. Rev. Lett.}

\PreprintCoverAbstract{\input{00_Abstract}}

\begin{document}

\title{\protect\input{00_Title.tex}}

\author{G. Aad \textit{et al.}}
\collaboration{ATLAS Collaboration}

\begin{abstract}
\protect\input{00_Abstract}
\end{abstract}


\maketitle

\label{sec:Introduction}
\input{01_Introduction.tex}

\label{sec:Data}
\input{02_Data.tex}

\label{sec:EventSelection}
\input{03_EventSelection}

\label{sec:SignalAcceptance}
\input{04_SignalAcceptance.tex}

\label{sec:Background}
\input{05_Background.tex}

\label{sec:Results}
\input{06_Results.tex}

\label{sec:Conclusion}
\input{07_Conclusions.tex}

\begin{acknowledgments}
\input{Acknowledgements.tex}
\end{acknowledgments}

\bibliographystyle{atlasBibStyleWoTitle}
\bibliography{Z4Leptons}{}

\onecolumngrid
\newpage
\input{atlas_authlist.tex}

\end{document}

%% file: Numbers.tex
\newcommand\LumiSevenTeV{\ensuremath{4.5\text{ fb}^{-1}}}
\newcommand\LumiEightTeV{\ensuremath{20.3\text{ fb}^{-1}}}
\newcommand\LumiUncertaintySevenTeV{\ensuremath{1.8\%}}
\newcommand\LumiUncertaintyEightTeV{\ensuremath{2.8\%}}

\newcommand\ZmumuBR{\ensuremath{\left( 3.366 \pm 0.007 \right)\%}}

\newcommand\AzmumuSevenTeV{\ensuremath{\left( 41.4 \pm 0.6 \right)\%}}
\newcommand\AzmumuEightTeV{\ensuremath{\left( 41.8 \pm 0.6 \right)\%}}

\newcommand\ObsEEEESevenTeV{\ensuremath{1}}
\newcommand\ObsEEMMSevenTeV{\ensuremath{7}}
\newcommand\ObsMMEESevenTeV{\ensuremath{5}}
\newcommand\ObsMMMMSevenTeV{\ensuremath{8}}
\newcommand\ObsAllSevenTeV{\ensuremath{21}}
\newcommand\ObsEEEEEightTeV{\ensuremath{16}}
\newcommand\ObsEEMMEightTeV{\ensuremath{48}}
\newcommand\ObsMMEEEightTeV{\ensuremath{16}}
\newcommand\ObsMMMMEightTeV{\ensuremath{71}}
\newcommand\ObsAllEightTeV{\ensuremath{151}}

\newcommand\ExpectedSigAllSevenTeV{\ensuremath{23.8 \pm 1.2}}

\newcommand\ExpectedSigAllEightTeV{\ensuremath{145 \pm 7}}

\newcommand\ExpBkgEEEESevenTeV{\ensuremath{0.12 \pm 0.04}}
\newcommand\ExpBkgEEMMSevenTeV{\ensuremath{0.18 \pm 0.09}}
\newcommand\ExpBkgMMEESevenTeV{\ensuremath{0.07 \pm 0.04}}
\newcommand\ExpBkgMMMMSevenTeV{\ensuremath{0.08 \pm 0.04}}
\newcommand\ExpBkgAllSevenTeV{\ensuremath{0.44 \pm 0.14}}
\newcommand\ExpBkgEEEEEightTeV{\ensuremath{0.14 \pm 0.03}}
\newcommand\ExpBkgEEMMEightTeV{\ensuremath{0.32 \pm 0.05}}
\newcommand\ExpBkgMMEEEightTeV{\ensuremath{0.18 \pm 0.04}}
\newcommand\ExpBkgMMMMEightTeV{\ensuremath{0.34 \pm 0.05}}
\newcommand\ExpBkgAllEightTeV{\ensuremath{1.0 \pm 0.11}}

\newcommand\UCTotalEEEightTeVSummary{\ensuremath{9.4\%}}
\newcommand\UCTotalMMEightTeVSummary{\ensuremath{2.7\%}}
\newcommand\UCTotalMEEightTeVSummary{\ensuremath{6.2\%}}
\newcommand\UCTotalEMEightTeVSummary{\ensuremath{3.7\%}}

\newcommand\CzfourlEEEESevenTeV{\ensuremath{21.5\%}}
\newcommand\CzfourlEEMMSevenTeV{\ensuremath{49.0\%}}
\newcommand\CzfourlMMEESevenTeV{\ensuremath{36.3\%}}
\newcommand\CzfourlMMMMSevenTeV{\ensuremath{59.2\%}}
\newcommand\CzfourlEEEEEightTeV{\ensuremath{36.1\%}}
\newcommand\CzfourlEEMMEightTeV{\ensuremath{55.5\%}}
\newcommand\CzfourlMMEEEightTeV{\ensuremath{46.2\%}}
\newcommand\CzfourlMMMMEightTeV{\ensuremath{71.1\%}}

\newcommand\FiducialXsecObsEEEESevenTeV{\ensuremath{0.9^{+1.4}_{-0.7} \pm 0.14 \pm 0.02 }}
\newcommand\FiducialXsecObsEEMMSevenTeV{\ensuremath{3.1^{+1.4}_{-1.1} \pm 0.16 \pm 0.05 }}
\newcommand\FiducialXsecObsMMEESevenTeV{\ensuremath{3.0^{+1.6}_{-1.2} \pm 0.30 \pm 0.06 }}
\newcommand\FiducialXsecObsMMMMSevenTeV{\ensuremath{3.0^{+1.2}_{-0.9} \pm 0.07 \pm 0.05 }}
\newcommand\FiducialXsecObsEEEEEightTeV{\ensuremath{2.2^{+0.6}_{-0.5} \pm 0.20 \pm 0.06 }}
\newcommand\FiducialXsecObsEEMMEightTeV{\ensuremath{4.2^{+0.7}_{-0.6} \pm 0.16 \pm 0.12 }}
\newcommand\FiducialXsecObsMMEEEightTeV{\ensuremath{1.7^{+0.5}_{-0.4} \pm 0.10 \pm 0.04 }}
\newcommand\FiducialXsecObsMMMMEightTeV{\ensuremath{4.9^{+0.7}_{-0.6} \pm 0.13 \pm 0.14 }}

\newcommand\PhaseSpaceXsecObsSameSevenTeVPlain{\ensuremath{32 \pm 11 \pm 1.0 \pm 0.6}}
\newcommand\PhaseSpaceXsecObsMixSevenTeVPlain{\ensuremath{44 \pm 14 \pm 3.3 \pm 0.9}}
\newcommand\PhaseSpaceXsecObsSevenTeVPlain{\ensuremath{76 \pm 18 \pm 4 \pm 1.4}}
\newcommand\PhaseSpaceXsecObsMixEightTeVPlain{\ensuremath{52 \pm 7 \pm 2.4 \pm 1.5}}
\newcommand\PhaseSpaceXsecObsSameEightTeVPlain{\ensuremath{56 \pm 6 \pm 1.8 \pm 1.6}}
\newcommand\PhaseSpaceXsecObsEightTeVPlain{\ensuremath{107 \pm 9 \pm 4 \pm 3.0}}

\newcommand\AzfourlEEEESevenTeV{\ensuremath{7.5\%}}
\newcommand\AzfourlMMMMSevenTeV{\ensuremath{18.3\%}}
\newcommand\AzfourlEEMMSevenTeV{\ensuremath{15.8\%}}
\newcommand\AzfourlMMEESevenTeV{\ensuremath{8.8\%}}
\newcommand\AzfourlEEEEEightTeV{\ensuremath{7.3\%}}
\newcommand\AzfourlMMMMEightTeV{\ensuremath{17.8\%}}
\newcommand\AzfourlEEMMEightTeV{\ensuremath{14.8\%}}
\newcommand\AzfourlMMEEEightTeV{\ensuremath{7.9\%}}

\newcommand\PhaseSpaceXsecTheorySevenTeV{\ensuremath{90.0 \pm 2.1\text{ fb}}}

\newcommand\PhaseSpaceXsecTheoryEightTeV{\ensuremath{104.8 \pm 2.5\text{ fb}}}

\newcommand\AltObsBranching{\ensuremath{\left( 4.31 \pm 0.34\text{ (stat)} \pm 0.17\text{ (syst)} \right) \times 10^{-6}}}

\newcommand\ObsBranching{\ensuremath{\left( 3.20 \pm 0.25\text{ (stat)} \pm 0.13\text{ (syst)} \right) \times 10^{-6}}}

\newcommand\TChanFracSame{\ensuremath{\left( 3.35 \pm 0.02 \right)\%}}
\newcommand\TChanFracDiff{\ensuremath{\left( 3.90 \pm 0.02 \right)\%}}

\newcommand\BkgATotalTotalSRSevenTeV{\ensuremath{0.38 \pm 0.14}}

\newcommand\BkgATotalTotalSREightTeV{\ensuremath{0.49 \pm 0.10}}

\newcommand\ExpectedTotalEEEESevenTeV{\ensuremath{1.8 \pm 0.3}}
\newcommand\ExpectedTotalEEMMSevenTeV{\ensuremath{7.9 \pm 0.4}}
\newcommand\ExpectedTotalMMEESevenTeV{\ensuremath{3.3 \pm 0.3}}
\newcommand\ExpectedTotalMMMMSevenTeV{\ensuremath{11.3 \pm 0.5}}
\newcommand\ExpectedTotalAllSevenTeV{\ensuremath{24.2 \pm 1.2}}
\newcommand\ExpectedTotalEEEEEightTeV{\ensuremath{14.4 \pm 1.4}}
\newcommand\ExpectedTotalEEMMEightTeV{\ensuremath{43.2 \pm 2.1}}
\newcommand\ExpectedTotalMMEEEightTeV{\ensuremath{19.3 \pm 1.3}}
\newcommand\ExpectedTotalMMMMEightTeV{\ensuremath{68.8 \pm 2.7}}
\newcommand\ExpectedTotalAllEightTeV{\ensuremath{146 \pm 7}}

\newcommand\AltExpectedBranching{\ensuremath{\left( 4.50 \pm 0.01 \right) \times 10^{-6}}}
\newcommand\ExpectedBranching{\ensuremath{\left( 3.33 \pm 0.01 \right) \times 10^{-6}}}

\newcommand\CMSObsBranching{\ensuremath{\left( 4.2^{+0.9}_{-0.8}\text{ (stat)} \pm 0.2\text{ (syst)} \right) \times 10^{-6}}}

%% file: 00_Title.tex
Measurement of the 4$\ell$ Cross Section at the $Z$ Resonance and 
Determination of the Branching Fraction 
of
$Z\rightarrow 4\ell$
in $pp$ Collisions 
at $\sqrt s$ = 7~and 8~\TeV{} 
with ATLAS

%% file: 00_Abstract.tex
Measurements of four-lepton (4$\ell$, $\ell=e,\mu$)
production \xsecs{}
at the $Z$~resonance in $pp$ collisions at the LHC 
with the ATLAS detector are presented.
For
dilepton and four-lepton invariant
mass region
$m_{\ell^+\ell^-} > 5$~\GeV{} 
and $80 < m_{4\ell} < 100$~\GeV{},
the measured \xsecs{} are
$76 \pm 18 \text { (stat) } \pm 4 \text { (syst) } \pm 1.4 \text { (lumi) }$ fb and
$107 \pm 9 \text{ (stat) } \pm 4 \text{ (syst) } \pm 3.0 \text { (lumi) }$ fb
at $\sqrt s$ = 7 and 8~\TeV{}, respectively.
By subtracting the non-resonant 4$\ell$ production contributions and normalizing with $Z\rightarrow \mu^+\mu^-$ events, 
the branching fraction for the $Z$~boson decay to $4\ell$ is
determined to be \ObsBranching{},
consistent with the Standard Model prediction.

%% file: 01_Introduction.tex
\begin{figure}[hb]
  \begin{tabular}{c@{\hskip 5mm}c}
  \includegraphics[width=.22\textwidth]{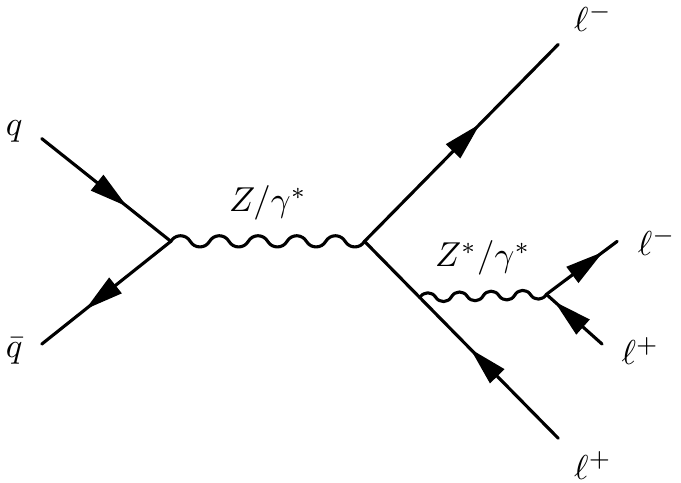}
  &
  \includegraphics[width=.22\textwidth]{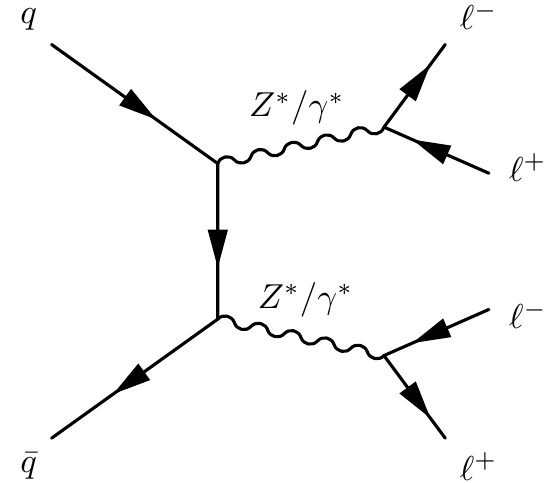}
  \\
  (a) & (b) \\
  \end{tabular}
  \caption{
    Examples of (a) \mbox{$s$-channel} and (b) \mbox{$t$-channel} Feynman diagrams for 4$\ell$ production in $pp$ collisions.
  }
  \label{fig:z4l_feynman}
\end{figure}

This Letter presents measurements of the \xsecs{} for the inclusive production
of four leptons (4$\ell$, $\ell = e, \ \mu$)   
at the $Z$ resonance in proton--proton collisions
at $\sqrt s$ = 7 and 8~\TeV{}
using data recorded by the ATLAS detector~\cite{bib:ATLASDetectorPaper}  
at the LHC~\cite{bib:LHCMachine} in 2011 and 2012.
In the Standard Model (SM), $4\ell$ production in the $Z$ resonance region occurs dominantly
via an \mbox{$s$-channel} diagram such as shown in Fig.~\ref{fig:z4l_feynman}(a)
where the $Z$ boson decay to charged leptons includes the production of an additional lepton pair from
the internal conversion of a virtual $Z$ or $\gamma$.
A small fraction of 4$\ell$ events is also produced in a \mbox{$t$-channel} process
such as shown in Fig.~\ref{fig:z4l_feynman}(b).
The process $gg\rightarrow Z^{(*)}Z^{(*)}\rightarrow 4\ell$
accounts for only about $10^{-3}$ of the total 4$\ell$ event rate
around the $Z$ resonance~\cite{Campbell:2011bn}.
A resonant peak around the $Z$ mass in the 4$\ell$ invariant mass
spectrum is observed along with the nearby peak from the Higgs boson
decay $H\rightarrow 4\ell$~\cite{Aad20121,Chatrchyan201230}.
A measurement of the 4$\ell$ production cross section at the $Z$ resonance 
provides a test of the SM
and a cross-check of the detector response to 
the 4$\ell$ final state from Higgs decays.

Since the interference between the 
resonant and non-resonant ($t$-channel and $gg$) production mechanisms
is expected to be very small around the $Z$ resonance,
the branching fraction of the rare decay $Z\rightarrow 4\ell$ can be determined
by subtracting the expected non-resonant 4$\ell$ contributions from the measured
4$\ell$ rate. 
For simplicity, inclusive 4$\ell$ production around the $Z$ resonance,
including the non-resonant contributions,
is denoted as $Z\rightarrow 4\ell$ from here on, 
except that the branching fraction $\ZllllFrac{}$ refers to the $s$-channel contribution alone.
The CMS Collaboration has observed the \SingleZllll{} resonance in $\sqrt{s}=7$~\TeV{} data
and determined a branching fraction, summed over the $4e$, $4\mu$ and $2e2\mu$ final states, of
$\ZllllFrac{} = \CMSObsBranching{}$,
where $80 < m_{4\ell} < 100$~\GeV{} and 
$m_{\ell\ell} > 4$~\GeV{} for all pairs of leptons  
(regardless of lepton flavor or charge)~\cite{Zto4lCMS}.
The results presented here include 
the first cross section measurement of the 4$\ell$ production at the $Z$ resonance 
at $\sqrt s$ = 8~\TeV{}, 
and a determination of $\ZllllFrac{}$ with improved statistical precision.

The inclusive 4$\ell$ production \xsec{} at the $Z$ resonance
is measured separately for the 4$e$, 4$\mu$ and 2$e$2$\mu$ final states
for each of the  $\sqrt s $ = 7 and 8~\TeV{} datasets
in a fiducial region (defined below) corresponding 
closely to the experimental acceptance.
The measured fiducial \xsecs{} are then extrapolated to 
a final phase-space region defined by the dilepton and four-lepton 
invariant mass requirements $m_{\ell^+\ell^-} > 5$~\GeV{} 
and $80 < m_{4\ell} < 100$~\GeV{},
where $\ell^+\ell^-$ denotes all same-flavor lepton pairs with opposite charge.
The branching fraction $\ZllllFrac{}$ is determined by normalizing the
resonant 4$\ell$ production rate to the
$Z \rightarrow \mu^+\mu^-$ production rate measured in the same dataset.

The ATLAS detector
has a cylindrical geometry~\footnote{ATLAS uses
a right-handed coordinate system with its origin
at the nominal interaction point(IP) in the center of the detector,
and the $z$-axis along the beam line. The $x$-axis points from the IP to
the center of the LHC ring, and the $y$-axis points upwards. Cylindrical coordinates
($r,\ \phi$) are used in the transverse plane, $\phi $ 
being the azimuthal angle around the beam line.
The pseudorapidity is defined
in terms of the polar angle $\theta $ as $\eta=-\ln\tan(\theta/2)$.
Observables labeled ``transverse'' are projected into the $x$--$y$ plane.}
and consists of an inner tracking detector surrounded by a 2~T superconducting solenoid, 
electromagnetic and hadronic calorimeters, and a muon spectrometer with a toroidal magnetic field. 
The inner detector (ID) provides precision tracking for charged particles for $|\eta|<2.5$. 
It consists of silicon pixel and strip detectors surrounded by a straw tube tracker that 
also provides transition radiation measurements for electron identification. The calorimeter 
system covers the pseudorapidity range $|\eta|< 4.9$. 
For $|\eta|<2.5$, the liquid-argon 
electromagnetic calorimeter is finely segmented and plays an important role 
in electron identification.
The muon spectrometer includes fast-trigger chambers ($|\eta|<2.4$) 
and high-precision tracking chambers covering $|\eta|<2.7$.

%% file: 02_Data.tex
The datasets for this analysis are recorded using single-lepton and dilepton triggers.
The transverse momentum (\pt{}) thresholds of these triggers vary from 20 to 24~\GeV{}
for the single-lepton triggers and from 8 to 13~\GeV{} for the dilepton triggers,
depending on lepton flavor and data-taking period.
The overall trigger efficiency for \Zllll{} events ranges from
94 to 99\%.

After removing the short data-taking periods having problems that affect the lepton reconstruction,
the total integrated luminosity used in the analysis is
\LumiSevenTeV{} at 7~\TeV{} and \LumiEightTeV{} at 8~\TeV{}.
The overall uncertainty on the integrated luminosity 
is $\LumiUncertaintySevenTeV{}$~\cite{lumi7tev}
and $\LumiUncertaintyEightTeV{}$~\footnote{
The 2012 luminosity measurement follows the same methodology as that detailed in Ref.~\cite{lumi7tev}. It is based
on a preliminary calibration of the luminosity scale derived from beam-separation scans performed in November 2012.
}
for the $\sqrt s$ = 7 and 8~\TeV{} datasets, respectively. 

The \progname{Powheg} Monte Carlo (MC) program~\cite{Melia:2011tj,Alioli:2010xd,Nason:2004rx},
used to calculate the signal cross sections,
includes perturbative QCD corrections to next-to-leading order.
The calculation also includes the interference terms
between the \mbox{$s$-channel} and the \mbox{$t$-channel} as well as
the interference terms between the $Z$ and the $\gamma^*$ diagrams.
The CT10~\cite{Lai:2010vv} set of parton distribution functions (PDFs) and 
QCD renormalization and factorization
scales of $\mu_\mathrm{R}, \mu_\mathrm{F} = m_{4\ell}$ are used.
In the $m_{\ell^+\ell^-} > 5$~\GeV{} and $80 < m_{4\ell} < 100$~\GeV{} phase space,
the production \xsecs{} calculated by \progname{Powheg} are
$53.4\pm 1.2$ fb ($45.8\pm 1.1$ fb) for the sum of the $4e$ and $4\mu$ final states, and
$51.5\pm 1.2$ fb ($44.2\pm 1.1$ fb) for the $2e2\mu$ final state at 8~\TeV{} (7~\TeV{}).
The cross sections for $4e$ and $4\mu$ are larger than for $2e2\mu$ due to
the interference between the two same-flavor lepton pairs.
The expected fraction of 4$\ell$ events produced via the \mbox{$t$-channel} process, $f_t$, is
\TChanFracSame{} and \TChanFracDiff{}
for same-flavor ($4e$, $4\mu$) and mixed-flavor ($2e2\mu$) final states, respectively,
for both 7 and 8~\TeV{}.
The $gg\rightarrow ZZ\rightarrow 4\ell$ process is
modeled by \progname{gg2zz}~\cite{Binoth:2008pr}, 
and the 4$\ell$ event fraction from this process
is calculated to be around 0.1\%.
The overall non-resonant fraction ($f_{\mathrm{nr}}$)
from the $t$-channel and $gg$ contributions combined is
($3.45\pm 0.02$)\% and ($4.00\pm 0.02$)\%
for the same-flavor and mixed-flavor final states, respectively. 
To generate MC events with a simulation of the detector to determine the signal acceptance,
\progname{Powheg} is interfaced to \progname{Pythia6}~\cite{pythia} 
or \progname{Pythia8}~\cite{pythia8} for showering and hadronization
and to \progname{Photos}~\cite{Golonka:2005pn} for radiated photons from charged leptons.

The MC generators used to simulate the reducible background contributions
are \progname{MC@NLO}~\cite{bib:mcatnlo} (to model $t\bar t$ and single-top production)
and \progname{Alpgen}~\cite{Mangano:2002ea} (to model \Z{}~boson 
production in association with jets, referred to as $Z$+jets).
These generators are interfaced to \progname{Herwig}~\cite{herwig} and \progname{Jimmy}~\cite{Jimmy}
for parton showering and underlying-event simulations.
The diboson background processes 
\WZ{} and $Z\gamma$, and $Z^{(*)}Z^{(*)}\rightarrow 4\ell $ decays involving $\tau\rightarrow e/\mu+2\nu$,
are modeled by
\progname{Powheg} (interfaced to \progname{Pythia} for parton showering) and
\progname{Sherpa}~\cite{Gleisberg:2008ta}.

The detector response simulation~\cite{simuAtlas} is based on the GEANT4 program~\cite{bib:geant4a}.
Additional inelastic $pp$ interactions (referred to as pile-up)  are included in the simulation, and
events are re-weighted to reproduce the observed distribution of the average number of collisions per
bunch-crossing in the data.

%% file: 03_EventSelection.tex
The \Zllll{} event selection closely follows the 
$H \rightarrow ZZ^* \rightarrow 4\ell$ analysis~\cite{Aad2012-higgs}
with muon \pt{}
and dilepton invariant mass requirements loosened
to increase the acceptance for the \Zllll{} process.

Muons are identified by tracks (or track segments) reconstructed in the muon spectrometer
and are matched to tracks reconstructed in the ID.
The muon momentum is calculated by combining the information from the two subsystems,
correcting for the energy lost in the calorimeters.
Additionally, one muon in each event is allowed to be a 
stand-alone muon 
or a calorimeter-tagged muon, 
where the stand-alone muon is 
identified by only a muon spectrometer track in $2.5<|\eta|<2.7$,
and the calorimeter-tagged muon is identified by an ID track with \pt{} $>$ 15~\GeV{}
associated with a compatible calorimeter energy deposit in $|\eta|<0.1$.
All muon candidates are required to have \pt{} $>$ 4~\GeV{} and $|\eta|<2.7$.

Electrons are reconstructed from energy deposits in the electromagnetic calorimeter
matched to a track in the ID~\cite{egammanote}.  
Tracks associated with electromagnetic clusters are fitted using a Gaussian Sum Filter~\cite{egammaGSF},
which allows bremsstrahlung energy losses to be taken into account.
For $\sqrt{s}=8$~\TeV{} data, improved 
electron discrimination from jets is obtained using a likelihood function formed from
parameters characterizing the shower shape and track association,
resulting in a reduction of the electron misidentification rate by more than a factor of two
compared to that at 7 TeV.
Electron candidates are required to have \pt{} $>$ 7~\GeV{} and $|\eta| < 2.47$.

Collision events are selected by requiring at least one reconstructed vertex
with at least three charged particle tracks with \pt{} $>$ 0.4~\GeV{}.
If more than one vertex satisfies the selection requirement,
the primary vertex is chosen as the one with the highest  $\sum p_{\mathrm{T}}^2$, 
summed over all tracks associated with the vertex.

In order to reject electrons and muons from jets, only isolated leptons are selected, requiring
the scalar sum of the transverse momenta, $\sum p_{\mathrm{T}}$, of other tracks inside a cone size of
$\Delta R = \sqrt{(\Delta\eta)^2 + (\Delta\phi)^2} = 0.2$ around the lepton to be less than 15\%
of the lepton \pt{}.
In addition, the $\sum \ET$ deposited in calorimeter cells inside a cone size of
$\Delta R = 0.2$ around the lepton direction, excluding the transverse energy due to the lepton and
corrected for the expected pile-up contribution, 
is required to be less than 30\% of the lepton \pt{}, reduced to
20\% for electrons in the 8~\TeV{} dataset and 15\% for stand-alone muons.
To further reject leptons from heavy-flavor jets,
the impact parameter relative to the primary vertex
is required to be less than
3.5 (6.0) standard deviations for all muons (electrons),
where the looser electron requirement allows for tails in the electron impact parameter distribution due to bremsstrahlung in the ID.
\begin{figure*}
\begin{tabular}{ccc}
\includegraphics[width=.3\linewidth]{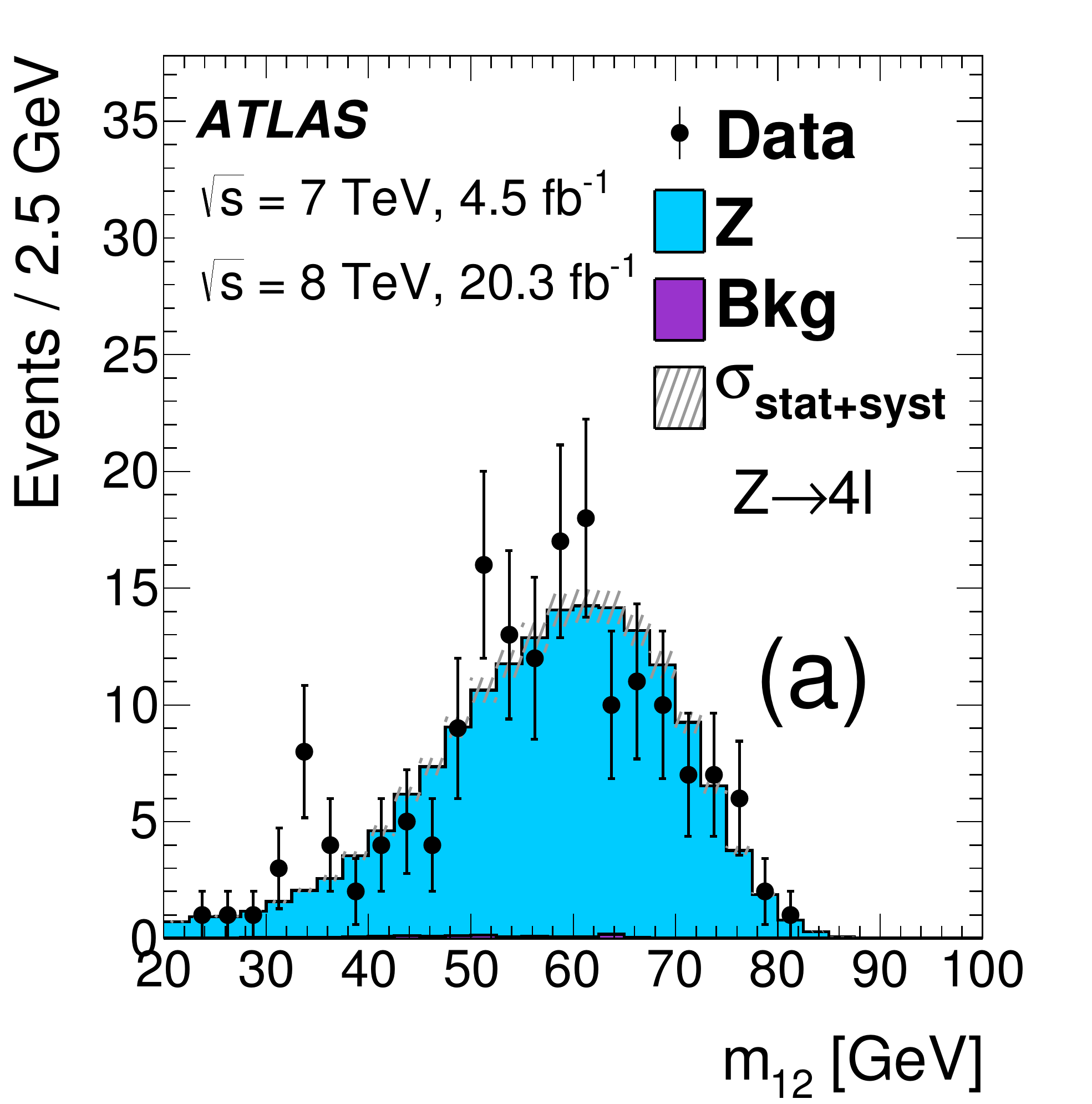} &
\includegraphics[width=.3\linewidth]{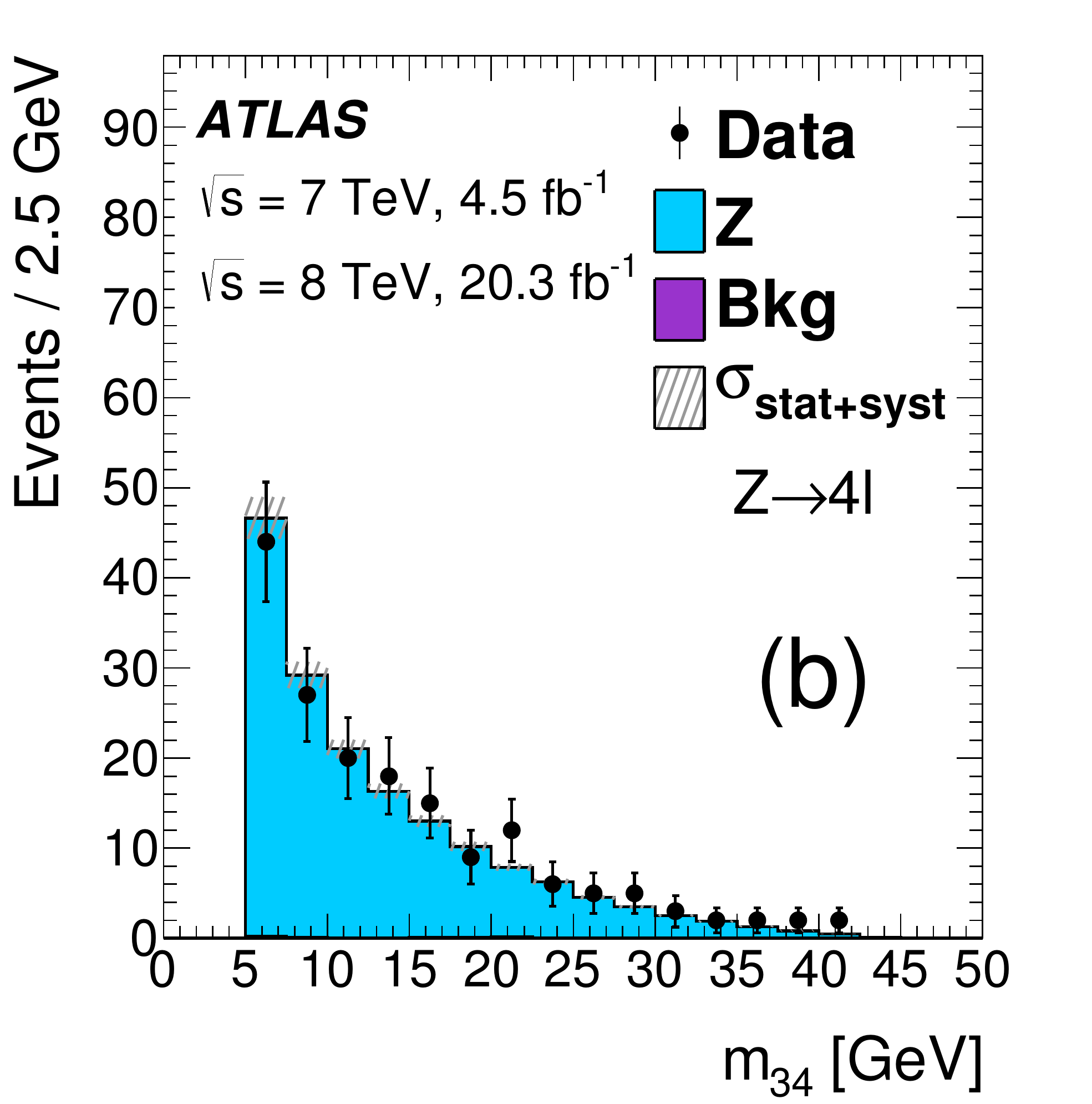} &
\includegraphics[width=.3\linewidth]{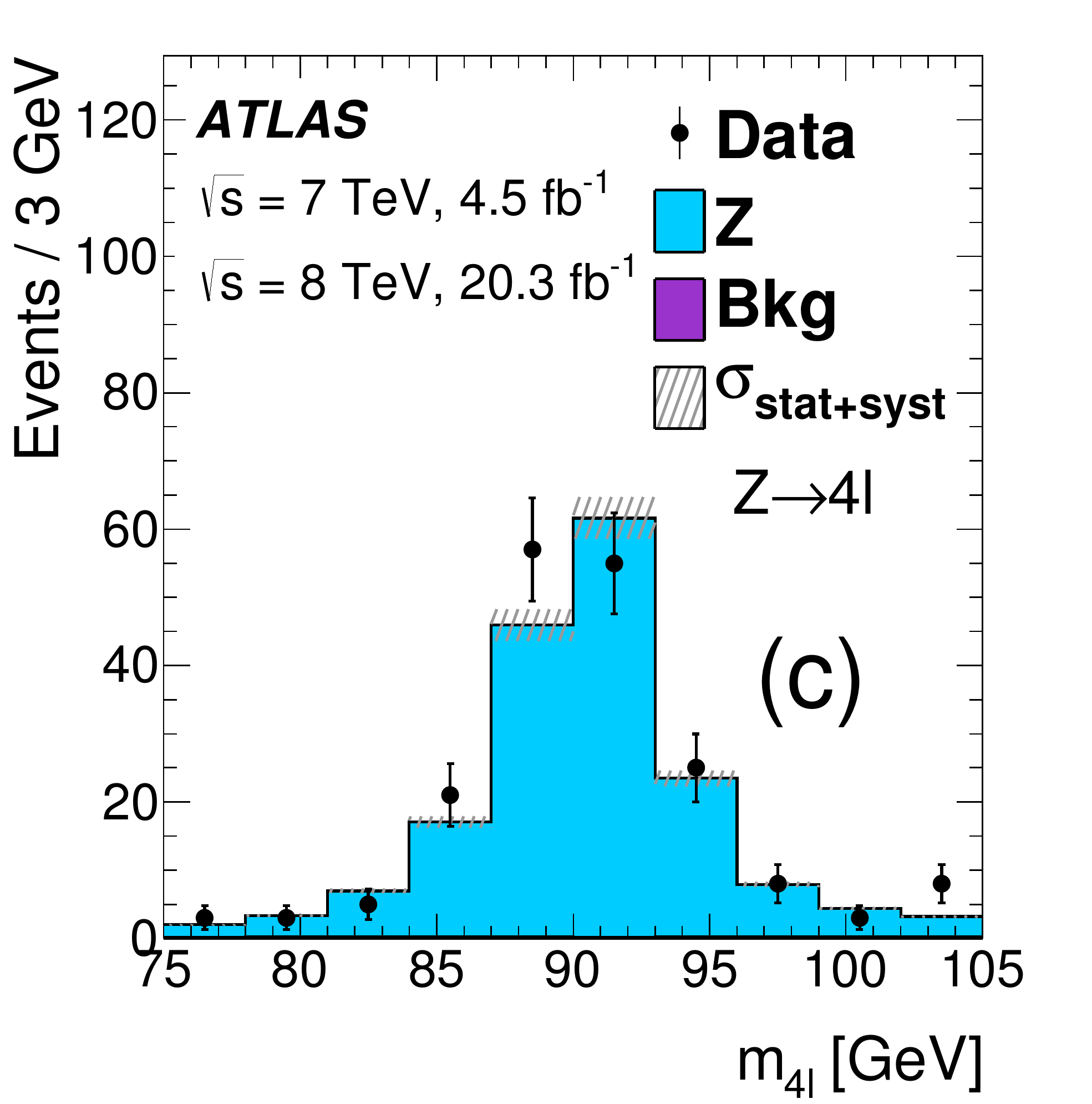} \\
\end{tabular}
\caption{
Invariant mass distributions of (a) the leading lepton pair, $m_{12}$, (b) the subleading lepton pair, $m_{34}$ ,
and (c) the four-lepton system, $m_{4\ell}$.
The MC simulation expectation for a combination of all channels is compared to 
$\sqrt{s}$ = 7 and 8~\TeV{} data.
All selections are applied except in (c) there is no $m_{4\ell}$ requirement.
The background contributes $< 1\%$ of the total expected signal (invisible in the plots).
}
\label{fig:z4l_mass}
\end{figure*}

Candidate quadruplets are formed by
selecting two opposite-sign, same-flavor dilepton pairs in an event.
The four leptons of the quadruplets are required to be well separated,
$\Delta R >0.1$ for same-flavor lepton pairs and $\Delta R >0.2$ for $e \mu$ pairs.
The two leading leptons must have \pt{} $>$ 20~\GeV{} and \pt{} $>$ 15~\GeV{}.
The third lepton must have \pt{} $>$ 10 (8)~\GeV{} if it is an electron (muon).
The dilepton pair with greatest invariant mass (denoted $m_{12}$) is called the leading lepton pair,
while the sub-leading lepton pair is chosen to have the largest
invariant mass (denoted $m_{34}$) among the remaining possible pairs.
The dilepton masses must satisfy $m_{12} > 20$~\GeV{} and $m_{34} > 5$~\GeV{}.
In the $4e$ and $4\mu$ channels
all same-flavor, opposite-sign lepton pairs 
are required to have $m_{\ell^+\ell^-} > 5$~\GeV{}.
This helps to reject events containing $J/\psi\rightarrow \ell^+\ell^-$ decays.
The invariant mass
of the lepton quadruplet is restricted to $80 < m_{4\ell} < 100$~\GeV{}.
A total of \ObsAllSevenTeV{} and \ObsAllEightTeV{}
\Zllll{} candidate events are selected in the 7 and 8~\TeV{} datasets, respectively.
The distributions of $m_{12}$, $m_{34}$, and $m_{4\ell}$ are shown in Fig.~\ref{fig:z4l_mass}.
The number of events observed in each channel is shown in Table~\ref{tab:xsec}.
The $2e2\mu$ channel is divided into $ee+ \mu\mu$ and $\mu\mu + ee$ channels
indicating the lepton flavors of the leading and sub-leading lepton pairs.

\begin{table*}
  \caption{Summary of
    the observed (\NObs{}) and expected (\NExp{}) 
    number of selected \Zllll{} candidate events,
    and the estimated number of background events (\NBkg{})
    in each 4$\ell$ channel
    for $\sqrt{s}$ = 7 and 8~\TeV{}. The associated uncertainties are statistical and systematic combined.
    The central values of the acceptance and efficiency factors (\Allll{}) and (\Cllll{}),
    the measured fiducial \xsecs{} (\SigFid{}),
    and the total \xsecs{} for $m_{\ell^+\ell^-} > 5$~\GeV{}, $80 < m_{4\ell} < 100$~\GeV{} (\SigPS{})
    are also presented.
    The fiducial regions are defined in the text and are different for each channel.
    The \SigPS{} are given for same-flavor ($4e$ and $4\mu$), different-flavor ($2e2\mu$), 
    and all channels combined. 
    The uncertainties on \SigFid{} and \SigPS{} are the statistical 
    and systematic uncertainties, and the uncertainty due to the luminosity measurement.
  }
  \label{tab:xsec}
  \begin{ruledtabular}
  \newcommand{\twrw}[1]{\multirow{2}{*}{#1}}
  \begin{tabular}{llcccccclc}
    $\sqrt{s}$ & $4\ell$ state   & \NObs{}            & \NExp{}                      & \NBkg{}               & \Cllll{}               & \SigFid{} [fb]                      & \Allll{}               &                           & \SigPS{} [fb]                                \\
    \hline
    7~\TeV{}   & $ee+ee$         & \ObsEEEESevenTeV{} & \ExpectedTotalEEEESevenTeV{} & \ExpBkgEEEESevenTeV{} & \CzfourlEEEESevenTeV{} & \FiducialXsecObsEEEESevenTeV{} & \AzfourlEEEESevenTeV{} & \twrw{$\Big\}\; 4e,4\mu$} & \twrw{\PhaseSpaceXsecObsSameSevenTeVPlain{}} \\
               & $\mu\mu+\mu\mu$ & \ObsMMMMSevenTeV{} & \ExpectedTotalMMMMSevenTeV{} & \ExpBkgMMMMSevenTeV{} & \CzfourlMMMMSevenTeV{} & \FiducialXsecObsMMMMSevenTeV{} & \AzfourlMMMMSevenTeV{} &                           &                                              \\
               & $ee+\mu\mu$     & \ObsEEMMSevenTeV{} & \ExpectedTotalEEMMSevenTeV{} & \ExpBkgEEMMSevenTeV{} & \CzfourlEEMMSevenTeV{} & \FiducialXsecObsEEMMSevenTeV{} & \AzfourlEEMMSevenTeV{} & \twrw{$\Big\}\; 2e2\mu$}  & \twrw{\PhaseSpaceXsecObsMixSevenTeVPlain{}}  \\
               & $\mu\mu+ee$     & \ObsMMEESevenTeV{} & \ExpectedTotalMMEESevenTeV{} & \ExpBkgMMEESevenTeV{} & \CzfourlMMEESevenTeV{} & \FiducialXsecObsMMEESevenTeV{} & \AzfourlMMEESevenTeV{} &                           &                                              \\
               & combined        & \ObsAllSevenTeV{}  & \ExpectedTotalAllSevenTeV{}  & \ExpBkgAllSevenTeV{}  &                        &                                     &                        &                           & \PhaseSpaceXsecObsSevenTeVPlain{}            \\
  \hline
    8~\TeV{}   & $ee+ee$         & \ObsEEEEEightTeV{} & \ExpectedTotalEEEEEightTeV{} & \ExpBkgEEEEEightTeV{} & \CzfourlEEEEEightTeV{} & \FiducialXsecObsEEEEEightTeV{} & \AzfourlEEEEEightTeV{} & \twrw{$\Big\}\; 4e,4\mu$} & \twrw{\PhaseSpaceXsecObsSameEightTeVPlain{}} \\
               & $\mu\mu+\mu\mu$ & \ObsMMMMEightTeV{} & \ExpectedTotalMMMMEightTeV{} & \ExpBkgMMMMEightTeV{} & \CzfourlMMMMEightTeV{} & \FiducialXsecObsMMMMEightTeV{} & \AzfourlMMMMEightTeV{} &                           &                                              \\
               & $ee+\mu\mu$     & \ObsEEMMEightTeV{} & \ExpectedTotalEEMMEightTeV{} & \ExpBkgEEMMEightTeV{} & \CzfourlEEMMEightTeV{} & \FiducialXsecObsEEMMEightTeV{} & \AzfourlEEMMEightTeV{} & \twrw{$\Big\}\; 2e2\mu$}  & \twrw{\PhaseSpaceXsecObsMixEightTeVPlain{}}  \\
               & $\mu\mu+ee$     & \ObsMMEEEightTeV{} & \ExpectedTotalMMEEEightTeV{} & \ExpBkgMMEEEightTeV{} & \CzfourlMMEEEightTeV{} & \FiducialXsecObsMMEEEightTeV{} & \AzfourlMMEEEightTeV{} &                           &                                              \\
               & combined        & \ObsAllEightTeV{}  & \ExpectedTotalAllEightTeV{}  & \ExpBkgAllEightTeV{}  &                        &                                     &                        &                           & \PhaseSpaceXsecObsEightTeVPlain{}            \\
  \end{tabular}
  \end{ruledtabular}
\end{table*}

%% file: 04_SignalAcceptance.tex
The overall signal selection efficiency 
is the product of efficiency and acceptance factors, \Cllll{} and \Allll{}, respectively.
The efficiency factor \Cllll{} is the ratio of
the number of \Zllll{} events passing the reconstructed event selections
to the number in the fiducial region,
and is determined using the signal MC samples after the detector simulation.
The fiducial region, defined at the MC generator level 
using the lepton four-momenta, requires
$p_\mathrm{T} > 20, 15, 10(8), 7(4)$~\GeV{} 
and $|\eta| < 2.5(2.7)$ of the $p_\mathrm{T}$-ordered $e(\mu)$,
$\Delta R(\ell, \ell') > 0.1(0.2)$ for all same(different)-flavor lepton pairs,
$m_{\ell^+\ell^-} > 20$~\GeV{} for at least one lepton pair,
$m_{\ell^+\ell^-} >  5$~\GeV{} for all same-flavor lepton pairs,
and $80 < m_{4\ell} < 100$~\GeV{}.
The four-momenta of all final state photons
within $\Delta R = 0.1$ of a lepton are summed into the four-momentum of that lepton.
The acceptance factor \Allll{} is the fraction
of \Zllll{} events in the final phase space 
which falls into the fiducial region.
The \Cllll{} uncertainty is mostly experimental
and the \Allll{} uncertainty is entirely theoretical.
The \Allll{} and \Cllll{} values
are listed in Table~\ref{tab:xsec} for each channel and dataset.
The \Cllll{} values for 8~\TeV{} are larger than for 7~\TeV{} due to a variety of factors,
including electron identification improvements with better bremsstrahlung treatment 
and additional muon detector coverage.

The MC lepton identification and trigger efficiencies are corrected based on studies performed in data control regions.
The energy and momentum scales and resolutions
of the MC events are calibrated to reproduce data
from $Z\rightarrow \ell^+\ell^-$ and $J/\psi \rightarrow \ell^+\ell^-$ decays.
The uncertainties on the \Zllll{} signal detection efficiency
are determined by varying the nominal calibrations
(including lepton energy and momentum resolutions and scales,
and the trigger, reconstruction and identification efficiencies)
in the MC samples by one standard deviation.
For the 8~\TeV{} (7~\TeV{}) analysis,
the relative uncertainties on the $C_{4\ell}$ factors are
\UCTotalMMEightTeVSummary{} (2.7\%),     
\UCTotalEMEightTeVSummary{} (4.9\%),
\UCTotalMEEightTeVSummary{} (9.8\%),
and \UCTotalEEEightTeVSummary{} (14.9\%)
for $\mu\mu+\mu\mu$, $ee+\mu\mu$, $\mu\mu+ee$, and $ee+ee$, respectively,
where the labeling $\ell\ell+\ell'\ell'$ indicates the leading and sub-leading lepton pairs.
The major uncertainty contributions come from the lepton reconstruction 
and identification efficiencies.

The relative uncertainties on the $A_{4\ell}$ factors, 
evaluated using \progname{Powheg} MC samples,
range from 1.3\% to 1.7\% depending on channel.
The theoretical uncertainties reflect uncertainties
from the choice of QCD scales and PDFs.
The scales are varied independently from 0.5 to 2.0 times 
the nominal $\mu_\mathrm{R}, \mu_\mathrm{F} = m_{4\ell}$.
The PDF uncertainties are estimated by 
taking the sum in quadrature of the deviations of $A_{4\ell}$ for each
PDF error set (52 CT10 eigenvectors varied by one standard deviation)
and for an alternative PDF set, MSTW2008~\cite{bib:MSTW2008},
with respect to the nominal one.

%% file: 05_Background.tex
The overall background 
in the selected 4$\ell$ event sample is estimated to be below 1\%,
as shown in Table~\ref{tab:xsec}.
The background contributions 
from diboson production are estimated, using MC simulations,
to be $0.06 \pm 0.01$ and $0.49 \pm 0.04$ events
in the 7 and 8~\TeV{} datasets, respectively.
Background contributions from $Z$+jets and top production processes
are estimated from data. 
Such background events
may contain two isolated leptons from $\Z$ decays or from $\W$ decays in top events,
together with additional activity such as heavy-flavor jets
or misidentified components of jets yielding reconstructed leptons.
These backgrounds are estimated from data using a background-enriched control 
sample of $\ell\ell j_\ell j_\ell$ events,
selected with the standard signal requirements
except that lepton-like jets, $j_\ell$, are selected in place of two of the signal leptons.
Electron-like jets, $j_e$,  in the $\ell\ell j_\ell j_\ell$ control sample 
are obtained from electromagnetic clusters matched
to tracks in the ID that do not satisfy the identification criteria or isolation requirements.
Muon-like jets, $j_\mu$, are defined as muon candidates that fail the requirements
on isolation.
The reducible background in the signal sample is estimated by scaling each event
in the $\ell\ell j_\ell j_\ell$ control sample by $f_1\times f_2$, 
where the factor 
$f_i$ ($i=1,2$)
for each of the two lepton-like jets
depends on lepton flavor and \pt{}.
The factor $f$ is the ratio of the probability for a jet to
satisfy the signal lepton selection criteria
to the probability for the jet to satisfy the lepton-like jet criteria,
and is obtained from independent jet-enriched 
data samples dominated by $Z$+jets or $t\bar t$ events.
The uncertainties on $f$ are determined from the variations of $f$ in data samples obtained
with alternative lepton-like jet selections and
different jet compositions.
The estimated background from $Z$+jets and top processes
ranges from 0.05 to 0.20 events for the different channels and datasets;
for all 4$\ell$ channels combined it is estimated 
to be \BkgATotalTotalSRSevenTeV{} and \BkgATotalTotalSREightTeV{} events
for the 7 and 8~\TeV{} data, respectively.

%% file: 06_Results.tex
The numbers of signal events predicted by MC simulation are
\ExpectedSigAllSevenTeV{} and \ExpectedSigAllEightTeV{} for 7 and 8~\TeV{}, respectively.
The data and MC predictions, as shown in Fig.~\ref{fig:z4l_mass}, are in good agreement.
Denoting the integrated luminosity by $L$,
the measured fiducial \xsecs{} (\SigFid{}), determined by 
(\NObs{} - \NBkg{})/($ L \times C_{4\ell}$), 
are given in Table~\ref{tab:xsec}.

The cross section in the final phase space for each channel
is calculated by \SigFid{}/\Allll{}.
The cross sections obtained for the $ee+ee$ and $\mu\mu+\mu\mu$ channels, 
and for the $2e+2\mu$ and $2\mu+2e$ channels,
are compatible within errors and are combined
using $2\times 2$ covariance matrices.
The total inclusive 4$\ell$ \xsec{} is a sum of the two combined \xsecs{}, 
and the uncertainty includes  correlations between the four channels.
The \xsecs{} listed in Table~\ref{tab:xsec} 
are consistent with the SM predictions given earlier.

The \SingleZllll{} branching fraction, \ZllllFrac{},
is determined by subtracting the non-resonant contributions 
to the selected events and 
normalizing the resulting yield to
the observed number of $Z\rightarrow \mu^+\mu^-$ events in the same dataset,
\[
  \ZllllFracDisp{}
     = \left(\ZmmFracDisp{}\right)
       \frac{\left(\NObs{} - \NBkg{}\right) \left(1-f_{\mathrm{nr}}\right) C_{2\mu} \cdot A_{2\mu}}{\left(\NObsMuMu{} - \NBkgMuMu{}\right) \Cllll{} \cdot \Allll{}}
       ,
\]
where $\ZmmFrac{} = \ZmumuBR{}$~\cite{PDG},
\NObsMuMu{} is around 1.7 million and 8.9 million
in the 7 and 8~\TeV{} datasets, respectively,
and \CAmumu{} is \AzmumuSevenTeV{} and \AzmumuEightTeV{},
respectively.
The background (\NBkgMuMu{}) is estimated to be around 0.3\% of the selected $Z\rightarrow\mu^+\mu^-$ events,
where the multijet and $W+$jets contributions are 
obtained using a data-driven method similar to that used for the \Zllll{} background estimation,
and other background contributions are estimated with simulated MC samples.
The branching fraction for \SingleZllll{}, summed over all $\ell=e,\mu$ final states,
is determined with both the 7 and 8~\TeV{} datasets.
The measured branching fractions for each dataset are consistent within uncertainties and are combined, giving
\[
  \ZllllFrac{} = \ObsBranching{}
\]
in the final phase-space region,
where the systematic uncertainty includes a contribution (about $0.2\%$) due to 
the interference between the \mbox{$s$-channel} and \mbox{$t$-channel} processes, 
calculated using \progname{CalcHep}~\cite{CalcHEP}.
The measured branching fraction is consistent with the SM prediction of
\ExpectedBranching{},
calculated using \progname{Powheg}. 
For a larger final phase-space region defined by
$m_{\ell^+\ell^-} > 4$~\GeV{} and $80 < m_{4\ell} < 100$~\GeV{},
similar to that used by CMS, 
the acceptance factors $A_{4\ell}$ and the non-resonant fractions $f_{\mathrm{nr}}$,
and their uncertainties, are also evaluated (leaving the fiducial region unchanged),  
and
the measured branching fraction becomes
$\ZllllFrac{} = \AltObsBranching{}$, compared with
an SM prediction of \AltExpectedBranching{}.
This result is consistent
with the CMS result measured with data
collected from $pp$ collisions at 7 TeV.

%% file: 07_Conclusions.tex
In summary, measurements of the \xsecs{} of 4$\ell$ production
at the $Z$~resonance in $pp$ collision
data collected by the ATLAS detector at the LHC have been presented.
The datasets analyzed correspond to an integrated luminosity of
\LumiSevenTeV{} and \LumiEightTeV{} 
at $\sqrt{s}$ = 7 and 8~\TeV{}, respectively.
The total \Zllll{} production \xsecs{}
in the phase-space region $m_{\ell^+\ell^-}>5$~\GeV{}
and $80 < m_{4\ell} < 100$~\GeV{}
are measured to be $\SigPS{}$ = 
$76\pm 18 \text { (stat) } \pm 4 \text { (syst) } \pm 1.4 \text { (lumi) }$ fb
at 7~\TeV{}
and 
$107\pm 9 \text{ (stat) } \pm 4 \text{ (syst) } \pm 3.0 \text { (lumi) }$ fb
at 8~\TeV{}, 
consistent with the SM predictions of 
\PhaseSpaceXsecTheorySevenTeV{} and \PhaseSpaceXsecTheoryEightTeV{}, respectively.
Normalizing to the measured $Z\rightarrow \mu^+ \mu^-$ production,
the $Z \rightarrow 4\ell$ branching fraction  
is determined to be \ObsBranching{},
consistent with the SM prediction of \ExpectedBranching{}.

%% file: Acknowledgements.tex
% Version 19-Feb-2014
We thank CERN for the very successful operation of the LHC, as well as the
support staff from our institutions without whom ATLAS could not be
operated efficiently.

We acknowledge the support of ANPCyT, Argentina; YerPhI, Armenia; ARC,
Australia; BMWF and FWF, Austria; ANAS, Azerbaijan; SSTC, Belarus; CNPq and FAPESP,
Brazil; NSERC, NRC and CFI, Canada; CERN; CONICYT, Chile; CAS, MOST and NSFC,
China; COLCIENCIAS, Colombia; MSMT CR, MPO CR and VSC CR, Czech Republic;
DNRF, DNSRC and Lundbeck Foundation, Denmark; EPLANET, ERC and NSRF, European Union;
IN2P3-CNRS, CEA-DSM/IRFU, France; GNSF, Georgia; BMBF, DFG, HGF, MPG and AvH
Foundation, Germany; GSRT and NSRF, Greece; ISF, MINERVA, GIF, I-CORE and Benoziyo Center,
Israel; INFN, Italy; MEXT and JSPS, Japan; CNRST, Morocco; FOM and NWO,
Netherlands; BRF and RCN, Norway; MNiSW and NCN, Poland; GRICES and FCT, Portugal; MNE/IFA, Romania; MES of Russia and ROSATOM, Russian Federation; JINR; MSTD,
Serbia; MSSR, Slovakia; ARRS and MIZ\v{S}, Slovenia; DST/NRF, South Africa;
MINECO, Spain; SRC and Wallenberg Foundation, Sweden; SER, SNSF and Cantons of
Bern and Geneva, Switzerland; NSC, Taiwan; TAEK, Turkey; STFC, the Royal
Society and Leverhulme Trust, United Kingdom; DOE and NSF, United States of
America.

The crucial computing support from all WLCG partners is acknowledged
gratefully, in particular from CERN and the ATLAS Tier-1 facilities at
TRIUMF (Canada), NDGF (Denmark, Norway, Sweden), CC-IN2P3 (France),
KIT/GridKA (Germany), INFN-CNAF (Italy), NL-T1 (Netherlands), PIC (Spain),
ASGC (Taiwan), RAL (UK) and BNL (USA) and in the Tier-2 facilities
worldwide.

%% file: atlas_authlist.tex
% ATLAS Collaboration author list
% Data extracted on 06-Mar-2014 for paper reference STDM-2013-08
\begin{flushleft}
{\Large The ATLAS Collaboration}

\bigskip

G.~Aad$^{\rm 84}$,
B.~Abbott$^{\rm 112}$,
J.~Abdallah$^{\rm 152}$,
S.~Abdel~Khalek$^{\rm 116}$,
O.~Abdinov$^{\rm 11}$,
R.~Aben$^{\rm 106}$,
B.~Abi$^{\rm 113}$,
M.~Abolins$^{\rm 89}$,
O.S.~AbouZeid$^{\rm 159}$,
H.~Abramowicz$^{\rm 154}$
% ...
\bigskip
\\
% ...
$^{11}$ Institute of Physics, Azerbaijan Academy of Sciences, Baku, Azerbaijan\\
% ...
$^{84}$ CPPM, Aix-Marseille Universit{\'e} and CNRS/IN2P3, Marseille, France\\
% ...
$^{89}$ Department of Physics and Astronomy, Michigan State University, East Lansing MI, United States of America\\
% ...
$^{106}$ Nikhef National Institute for Subatomic Physics and University of Amsterdam, Amsterdam, Netherlands\\
% ...
$^{112}$ Homer L. Dodge Department of Physics and Astronomy, University of Oklahoma, Norman OK, United States of America\\
$^{113}$ Department of Physics, Oklahoma State University, Stillwater OK, United States of America\\
% ...
$^{116}$ LAL, Universit{\'e} Paris-Sud and CNRS/IN2P3, Orsay, France\\
% ...
$^{152}$ Institute of Physics, Academia Sinica, Taipei, Taiwan\\
% ...
$^{154}$ Raymond and Beverly Sackler School of Physics and Astronomy, Tel Aviv University, Tel Aviv, Israel\\
% ...
$^{159}$ Department of Physics, University of Toronto, Toronto ON, Canada\\
% ...
\end{flushleft}